\documentclass[english,journal=jctcce,manuscript=article,layout=twocolumn]{achemso}

\usepackage{flafter}
\usepackage{amsmath}
\usepackage{amssymb}
\usepackage{nicefrac}
\usepackage{color}
\usepackage{graphicx}
\usepackage{siunitx}
\usepackage[dvipsnames]{xcolor}
\usepackage{array,booktabs}
\usepackage{mathrsfs}
\usepackage{threeparttable}
\usepackage{multirow}
\usepackage{adjustbox}

\usepackage[font=footnotesize,labelfont=bf,labelsep=period,width=0.45\textwidth]{caption}
\usepackage[font=footnotesize,labelfont=bf,labelsep=period]{subcaption}

\usepackage[capitalize]{cleveref}
\crefname{figure}{Figure}{Figures}
\crefname{table}{Table}{Tables}
\crefname{equation}{Equation}{Equations}
\crefname{section}{Section}{Sections}

\newcommand{\Eh}{E_\mathrm{h}}

\newcommand{\Rnlike}{[\mathrm{Rn}]}
\newcommand{\dfz}{\delta^{(0)}}          % frozen-orbital shift
\newcommand{\dsc}{\Delta\varepsilon}     % self-consistent shift
\newcommand{\iunit}{\mathrm{i}}          % imaginary unit
\newcommand{\lambdabar}{\mkern0.75mu\mathchar'26\mkern-9.75mu\lambda} % reduced Compton wavelength

\title{Self-Consistent Interplay of Breit and QED Effects in Heavy Elements}

\author{Tanner Culpitt}
\affiliation[University of Washington]
{Department of Chemistry, University of Washington, Seattle, WA 98195, USA}
\email{tculpitt@uw.edu}

\author{Shiv Upadhyay}

\author{Xiaosong Li}
\affiliation[University of Washington]
{Department of Chemistry, University of Washington, Seattle, WA 98195, USA}
\email{xsli@uw.edu}
\begin{document}

\maketitle

\twocolumn[
\begin{@twocolumnfalse}
\begin{abstract}
Quantum electrodynamic (QED) corrections improve the accuracy of four-component Dirac--Coulomb--Breit electronic structure theory. This work assesses the relative importance of the Breit gauge term and QED radiative corrections, as well as the differences between their variational and perturbative treatments. The Uehling vacuum polarization and Pyykk\"o--Zhao self-energy potentials, implemented alongside the full frequency-independent Breit interaction in the ChronusQ program, are evaluated for atoms and molecules with both frozen and relaxed orbitals. Across the closed-shell radon-like isoelectronic series from Rn to Np$^{7+}$ ($Z = 86$--93), the self-consistent gauge contribution reaches nearly half the magnitude of the Uehling shift for the core $s$ shells. When self-energy corrections are included, the gauge contribution exceeds the combined QED shift for most non-$s$ shells, while offsetting approximately 13\% of the total QED shift for the core $s$ shells, with little dependence on $Z$. Orbital relaxation in the variational treatment has minimal impact on the radiative shifts of the $s$ shells but substantially suppresses the gauge contribution toward the valence region. Consequently, perturbative treatments based on frozen orbitals tend to overestimate the gauge contribution to the outer $s$-shell energies. In TlBr, the gauge term reduces the equilibrium bond length by $0.05$~pm, compared with a reduction of $0.08$~pm from the combined QED corrections. These results demonstrate that the gauge contribution is comparable in magnitude to QED radiative corrections and should not be regarded as a negligible higher-order effect in accurate electronic structure calculations of either atoms or molecules.
\end{abstract}
\end{@twocolumnfalse}
]

%===============================================================================
\section{Introduction}
%===============================================================================

The leading corrections from quantum electrodynamics (QED) to the relativistic
electronic structure of atomic systems are the Breit
interaction\cite{Breit29_553,Gaunt29_151,Breit32_616} and the radiative
corrections, namely vacuum polarization\cite{Uehling35_55} and electron
self-energy.\cite{Faegri07_book,Grant07_book} Nearly three decades ago
valence-level Lamb shifts were estimated for the group 1 and 11
atoms.\cite{Pyykko98_R689} For valence properties of heavy elements it is
now well established that radiative corrections are small but not
negligible,\cite{Schwerdtfeger10_062503,Schwerdtfeger17_023002,Smits23_1} and
of a size comparable to the Breit
correction.\cite{Aucar18_134101}

A practical route toward the inclusion of radiative corrections for many-electron atoms and molecules is to represent them by effective one-electron operators.\cite{Liu15_631,Liu23_e1652} The leading vacuum polarization term is the Uehling potential, for which efficient evaluations
for finite nuclei are
available.\cite{FullertonRinker76_1283,Huang76_1311,Borie82_67} Higher-order contributions include
the Wichmann--Kroll\cite{Wichmann56_843} and K\"{a}ll\'{e}n--Sabry\cite{Sabry55_17} terms. The
self-energy has no exact local form and several effective potentials exist:
The single-Gaussian model of Pyykk\"{o} and
Zhao,\cite{PyykkoZhao03_1469} the potential of Flambaum and
Ginges,\cite{Flambaum05_052115} and the nonlocal model operator of Shabaev,
Tupitsyn, and Yerokhin\cite{Shabaev13_012513} with its extension to
superheavy elements.\cite{Malyshev22_012806} The first two are the ones in
common molecular use, and they are in relative agreement on $s$ orbitals but not
elsewhere. For example, on the $p$, $d$ and $f$ orbital energies of Au they differ substantially and often in sign,\cite{Saue22_164101,Janke25_104111} and
their QED corrections to parity-violating energies differ.\cite{Sunaga21_e1974592}

In atomic physics these potentials have long been combined with the Breit interaction in numerical (finite-difference or B-spline) bases. As early as 1976 the Gaunt, gauge-term, and
vacuum-polarization contributions were tabulated for subshells of
neutral atoms.\cite{Mark76_243} The Gaunt and gauge terms have also been
separated for superheavy ions,\cite{Indelicato86_1719} and multiconfiguration Dirac--Fock
calculations have progressively brought the Breit interaction and the
Uehling potential into the self-consistent
field.\cite{Indelicato90_5139,Indelicato92_2426,Boucard00_59,Rodrigues04_117,Indelicato07_155}
Self-consistent and perturbative Breit have been investigated at the orbital level
in Hg\cite{Oster89_2447} and, together with perturbative QED, in heavy
atoms.\cite{Schwerdtfeger10_062503} More recent tabulations over a wide range
of $Z$ calculated first-order QED alongside the full Breit correction,\cite{Aucar18_044113,Aucar18_134101}
while the gauge term has
been resolved orbital by orbital in a Gaussian basis, without radiative
corrections.\cite{Hoyer26_5851} For orbitals with $l > 0$ the radiative shift
is dominated by the response of the core and can change sign relative to the
first-order term,\cite{Johnson04_054502,Ginges16_052509,Ginges16_095001}
as can the vacuum-polarization correction to hyperfine
structure.\cite{Hasted25_032812} Self-consistent Breit and variational QED
appear together in correlated studies of actinium\cite{Dzuba19_022504} and
gold,\cite{Schwerdtfeger17_023002} and rigorous treatments of one-electron
systems continue to advance,\cite{Oreshkina20_032511} including Gaussian-basis
formulations of the one-loop self-energy and of the vacuum-polarization
density for hydrogen-like ions.\cite{Salman23_012808,Ferenc25_L040802,Benazzouk26_032813}

In molecules, effective QED entered through one-component scalar
Douglas--Kroll--Hess calculations with the Uehling and Pyykk\"{o}--Zhao
potentials\cite{Peterson05_10363,Peterson07_164304,Peterson15_074105,Peterson16_184309,Peterson17_084108}
and through QED-fitted
pseudopotentials,\cite{Hangele12_214105,Hangele13_174113,Hangele13_044104,Hangele14_222}
and has since been implemented in four-component Gaussian-basis
calculations\cite{Skripnikov21_201101,Skripnikov21_144103,Sunaga21_e1974592,Saue22_164101,Sunaga26_68}
and in two-component zeroth-order regular approximation (ZORA) and exact two-component (X2C) calculations.\cite{Janke25_104111,Aucar26_022809,Cheng26_3988} QED corrections
to NMR couplings have also been estimated by scaling hydrogen-like
data.\cite{ColomboJofre22_064103}

Many of these calculations have been carried out at the Dirac--Coulomb or Dirac--Coulomb--Gaunt level, with the Gaunt term either in the Hamiltonian or added as a separate increment.\cite{Saue22_164101,Skripnikov21_144103,Janke25_104111,Peterson26_2571687} The full Breit interaction has entered molecular calculations at several levels. Its contribution has been taken from atomic calculations and added to molecular results alongside the QED corrections,\cite{Cheng22_13587,Aucar26_022809} or included in the molecular Hamiltonian through atomic mean-field integrals in exact two-component methods.\cite{Cheng22_4537,Cheng22_151101,Cheng22_6732,Cheng25_031404} Self-consistent Dirac--Fock--Breit calculations extend to molecules at the four-component level,\cite{Shiozaki13_204113,Li22_064112} and for light molecules the Breit interaction has been treated variationally with explicitly correlated Gaussians.\cite{Ferenc22_094113}

Beyond the Gaunt level, the full Breit interaction and QED have been combined and studied in different ways. Zhang, Cheng, and co-workers have resolved the gauge term in the bond lengths of hydrogen halides, dihalogens and methyl halides,\cite{Cheng22_4537,Cheng22_151101} and Sunaga, Salman, and Saue,\cite{Saue22_164101} and Bonar and Peterson\cite{Peterson26_2571687} have compared QED effects on bond lengths with those of the Gaunt term. Flynn, Quiney, and Grant include the full Breit interaction self-consistently together with variational vacuum polarization for E119$^+$ and E119F.\cite{Flynn25_042810,Flynn25_042811} Wang and co-workers have compared gauge-term and QED contributions to the zero-field splittings of chalcogen diatomics, in two-component calculations where the Breit interaction enters through atomic mean-field integrals.\cite{Cheng26_3988} Each of these studies therefore addresses a different combination of terms: the gauge term without QED, QED combined with the Gaunt term, vacuum polarization with the undivided Breit interaction, and the gauge term compared with QED, with the Breit interaction approximated by atomic mean-field integrals. The present work brings these elements together in a single self-consistent four-component treatment.

In this work, we have implemented the Uehling and Pyykk\"{o}--Zhao potentials variationally in the ChronusQ
program,\cite{Li20_e1436,Li26_046102} inside the Pauli-spinor Dirac--Hartree--Fock
framework,\cite{Li21_3388,Li26_154102} which already carries the
full Breit operator.\cite{Li22_064112,Li23_171101,Li23_9009} The Gaunt or Breit interaction, the
Uehling potential, and the self-energy potential can therefore be switched on in
any combination, and each can be evaluated either on
perturbative frozen orbitals or self-consistently. We apply our implementation to the radon-like
isoelectronic series, neutral Rn through Np$^{7+}$. Every member has the same
closed-shell configuration, $\Rnlike = 1s^2 \ldots 5d^{10} 6s^2 6p^6$, 86
electrons in 24 relativistic subshells. Because only $Z$ changes, each subshell
can be followed as a function of nuclear charge, and every self-consistent
calculation converges cleanly to a single well-defined state. On this series
the Gaunt interaction, the Breit interaction, the Uehling potential, and the self-energy can be compared subshell by subshell, on both frozen and relaxed orbitals. In addition, we compute the equilibrium bond length of TlBr with several
combinations of the Gaunt, Breit, and radiative terms in order to understand
the relative impact of the different contributions in a molecular setting. 

%===============================================================================
\section{Theory}
%===============================================================================

\subsection{Hamiltonian}

The reference Hamiltonian is the four-component Dirac--Coulomb (DC) operator,
\begin{equation}\label{eq:dc}
 \hat{H}_\mathrm{DC} = \sum_i \hat{h}^\mathrm{D}(i)
 + \sum_{i<j} \frac{1}{r_{ij}},
\end{equation}
where $\hat{h}^\mathrm{D}$ is the one-electron Dirac operator,\cite{Faegri07_book} the indices $i$ and
$j$ run over electrons, and $r_{ij} = |\mathbf{r}_{ij}|$ is the distance
between electrons $i$ and $j$. Atomic units are used throughout, so that the
electron mass is $m = 1$ and $c$ is the speed of light in those units. To this
we add, in the frequency-independent ($\omega_{ij} \to 0$) limit, the full
Breit operator\cite{Breit29_553,Breit32_616}
\begin{equation}\label{eq:breit}
 \hat{B}_{ij} = -\frac{1}{2 r_{ij}}
 \left[ \boldsymbol{\alpha}_i \cdot \boldsymbol{\alpha}_j
 + \frac{(\boldsymbol{\alpha}_i \cdot \mathbf{r}_{ij})
         (\boldsymbol{\alpha}_j \cdot \mathbf{r}_{ij})}{r_{ij}^2} \right],
\end{equation}
which separates into the Gaunt (magnetic) interaction\cite{Gaunt29_151}
\begin{equation}\label{eq:gaunt}
 \hat{G}_{ij} = -\frac{\boldsymbol{\alpha}_i \cdot \boldsymbol{\alpha}_j}{r_{ij}}
\end{equation}
and the gauge term
\begin{equation}\label{eq:gauge}
 \hat{B}_{ij} - \hat{G}_{ij} =
 \frac{\boldsymbol{\alpha}_i \cdot \boldsymbol{\alpha}_j}{2 r_{ij}}
 - \frac{(\boldsymbol{\alpha}_i \cdot \mathbf{r}_{ij})
         (\boldsymbol{\alpha}_j \cdot \mathbf{r}_{ij})}{2 r_{ij}^3} .
\end{equation}
The gauge term is often referred to as retardation,\cite{Mann71_41,Indelicato86_1719,Aucar18_044113}
although it is instantaneous, being the difference
between the zero-frequency electron--electron interactions in the Coulomb gauge
(Breit) and the Feynman gauge
(Gaunt).\cite{Sucher88_L585,Desclaux87_639,Indelicato88_1087,MartenssonPendrill89_3794,Lindgren90_1085} Here
$\boldsymbol{\alpha}_i = (\alpha_{i,x}, \alpha_{i,y}, \alpha_{i,z})$ is the vector of
the Dirac matrices for electron $i$. These matrices are given by
\begin{equation}\label{eq:alpha}
 \alpha_{J} = \begin{pmatrix}
  \mathbf{0}_2 & \sigma_J \\
  \sigma_J & \mathbf{0}_2
 \end{pmatrix} , \qquad J = x, y, z ,
\end{equation}
and are built from the Pauli matrices $\sigma_J$,
\begin{equation}\label{eq:pauli}
\begin{split}
 \mathbf{I} &= \begin{pmatrix} 1 & 0 \\ 0 & 1 \end{pmatrix} , \qquad
 \sigma_x = \begin{pmatrix} 0 & 1 \\ 1 & 0 \end{pmatrix} , \\[4pt]
 \sigma_y &= \begin{pmatrix} 0 & -\iunit \\ \iunit & 0 \end{pmatrix} , \qquad
 \sigma_z = \begin{pmatrix} 1 & 0 \\ 0 & -1 \end{pmatrix} ,
\end{split}
\end{equation}
where $\mathbf{I}$ and $\mathbf{0}_2$ are the $2 \times 2$ identity and null
matrices.\cite{Li21_3388}

The gauge term is obtained as the difference
between a calculation with the Breit operator and one with the Gaunt operator. Both two-electron
operators are evaluated in
the Pauli spinor basis described in \cref{sec:pauli}.\cite{Li21_3388,Li22_064112}

\subsection{Matrix Representation in the Pauli Basis}
\label{sec:pauli}

Every operator above is evaluated in the Pauli representation,\cite{Li21_3388} on which the present
implementation is built. We
summarize only what is needed to establish where the QED potentials of
\cref{sec:qed} enter the matrix problem.

The one-electron Dirac equation may be written according to
\begin{equation}\label{eq:dirac4c}
\begin{pmatrix}
 \mathbf{I} \otimes V & c\,\boldsymbol{\sigma} \cdot \mathbf{p} \\[2pt]
 c\,\boldsymbol{\sigma} \cdot \mathbf{p} & \mathbf{I} \otimes (V - 2mc^2)
\end{pmatrix}
\begin{pmatrix} \psi^L \\[2pt] \psi^S \end{pmatrix}
= E \begin{pmatrix} \psi^L \\[2pt] \psi^S \end{pmatrix},
\end{equation}
where $\mathbf{p} = -\iunit\nabla$ is the momentum operator,
$\boldsymbol{\sigma} = (\sigma_x, \sigma_y, \sigma_z)$ are the Pauli matrices
of \cref{eq:pauli}, $\mathbf{I}$ is the $2 \times 2$ identity, $E$ is the
orbital energy, and $\psi^L$ and $\psi^S$ are the large- and small-component
two-spinors. The potential $V$ here is the nuclear attraction. The
small-component basis is generated from the large-component set through the
restricted kinetic balance condition,
\begin{equation}\label{eq:rkb}
 \chi^S_\mu = \frac{1}{2mc}\, \boldsymbol{\sigma} \cdot \mathbf{p}\, \chi^L_\mu ,
\end{equation}
in which $\chi^L_\mu$ is the $\mu$th large-component scalar basis function
and $\chi^S_\mu$ the small-component function generated from it. This
enforces the correct nonrelativistic limit of the positive-energy
states.\cite{Dyall90_25,Nieuwpoort91_131,Li21_3388}

Each $XY$ block of a four-component matrix, with $X, Y \in \{L, S\}$, is then
expanded in the Pauli basis, so that it carries one scalar component, labelled
$s$, and three spin-dependent components labelled $J = x, y, z$. In this
representation the core Hamiltonian matrix elements
are\cite{Li21_3388}
\begin{align}
 h^{LL}_{\mu\nu,s} &= 2 \langle \mu | V | \nu \rangle ,
 & h^{LL}_{\mu\nu,J} &= 0 , \label{eq:hLL} \\[2pt]
 h^{LS}_{\mu\nu,s} &= 2 \langle \mu | T | \nu \rangle ,
 & h^{LS}_{\mu\nu,J} &= 0 , \label{eq:hLS}
\end{align}
\begin{equation}\label{eq:hSSs}
 h^{SS}_{\mu\nu,s} = 2 \left[ \frac{1}{4m^2c^2}
 \nabla_\mu \!\cdot\! \nabla_\nu \langle \mu | V | \nu \rangle
 - \langle \mu | T | \nu \rangle \right] ,
\end{equation}
\begin{equation}\label{eq:hSSJ}
 h^{SS}_{\mu\nu,J} = \frac{2\iunit}{4m^2c^2}
 \left( \nabla_\mu \!\times\! \nabla_\nu \right)_J \langle \mu | V | \nu \rangle ,
\end{equation}
where $\mu$ and $\nu$ index the large-component scalar basis functions,
$T$ and $V$ are the kinetic and potential energy operators,
$\nabla_\mu$ denotes the gradient acting on $\chi^L_\mu$, and $J = x, y, z$
labels the three spin-dependent Pauli components. $\langle \mu | V | \nu
\rangle$ is shorthand for $\langle \chi^L_\mu | V | \chi^L_\nu \rangle$.

\Cref{eq:hLL,eq:hLS,eq:hSSs,eq:hSSJ} are the matrix elements through which
the QED potentials are implemented. Either potential of \cref{sec:qed} is
added to $V$. It therefore contributes to the large-component scalar block
through \cref{eq:hLL}, to the small-component scalar block through the
$\nabla_\mu \!\cdot\! \nabla_\nu$ term of \cref{eq:hSSs}, and through the $(\nabla_\mu \!\times\! \nabla_\nu)_J$ term of \cref{eq:hSSJ}
to the three spin-dependent parts of the small-component block.
\Cref{eq:hLS} is purely kinetic and is untouched.

\subsection{Effective QED Potentials}
\label{sec:qed}

The radiative corrections are represented by effective one-electron
potentials added to the nuclear attraction
\begin{equation}\label{eq:veff}
 V = V_\mathrm{nuc} + V_\mathrm{U} + V_\mathrm{SE},
\end{equation}
where $V_\mathrm{nuc}$ is the nuclear Coulomb potential,
$V_\mathrm{U}$ describes vacuum polarization, and $V_\mathrm{SE}$ the electron
self-energy. Vacuum polarization, self-energy, and the transverse part of single-photon exchange are all lowest-order
QED corrections to the energy of bound electrons.\cite{Saue22_164101}
The gauge term and the radiative potentials compared in this work therefore enter at the same order.
Both the self-energy and vacuum polarization potentials studied here are local,
spherically symmetric about the
nucleus and short-ranged. For further details surrounding the background of effective QED
potentials, we direct the reader to the work of Sunaga, Salman, and
Saue,\cite{Saue22_164101}  which is an excellent overview written for
a chemistry audience. In the following Sections dedicated to effective QED potentials, we
summarize only what is needed.

\subsection{Vacuum polarization: The Uehling Potential}
\label{sec:uehling}

In the field of the nucleus the vacuum is polarized: virtual
electron--positron pairs screen the nuclear charge, so that an electron
closer to the nucleus than roughly the reduced Compton wavelength of the electron, $\lambdabar$,
sees a larger
effective charge than one farther out. To leading order,
the resulting potential is the one derived by Uehling for a point
nucleus,\cite{Uehling35_55,FullertonRinker76_1283,Saue22_164101}
\begin{equation}\label{eq:uehling_point}
 V_\mathrm{U}^\mathrm{point}(r) = -\frac{2 Z \alpha}{3\pi r}\,
 K_1 \left( \frac{2r}{\lambdabar} \right) ,
\end{equation}
where $r$ is the distance from the nucleus, $\alpha = 1/c$ is the
fine-structure constant, and
\begin{equation}\label{eq:K1}
 K_1(x) = \int_1^\infty \mathrm{d}t\; e^{-xt}
 \left( \frac{1}{t^2} + \frac{1}{2t^4} \right) \sqrt{t^2 - 1} .
\end{equation}
The potential is attractive and adds to the nuclear attraction. For a finite
nucleus with charge density $\rho_\mathrm{n}$, normalized to $Z$, the
point-nucleus kernel is convolved with the charge
distribution to give,\cite{FullertonRinker76_1283,Saue22_164101}
\begin{equation}\label{eq:uehling_3D}
 V_\mathrm{U}(\mathbf{r}) = -\frac{2\alpha}{3\pi} \int \mathrm{d}^3 r' \,
 \frac{\rho_\mathrm{n}(\mathbf{r}')}{|\mathbf{r} - \mathbf{r}'|} \,
 K_1\!\left(\frac{2}{\lambdabar}|\mathbf{r} - \mathbf{r}'|\right),
\end{equation}
and the angular integration for a spherically symmetric nuclear charge distribution yields
\begin{equation}\label{eq:uehling_1D}
\begin{split}
V_\mathrm{U}(r) &= -\frac{2}{3}\frac{\alpha\lambdabar}{r} \int_0^\infty \mathrm{d}r' \,
r'\rho_\mathrm{n}(r') \,
\Bigg[ K_0 \left(\frac{2}{\lambdabar}|r - r'|\right) \\
&\qquad\qquad\qquad - K_0 \left(\frac{2}{\lambdabar}|r + r'|\right) \Bigg],
\end{split}
\end{equation}
where
\begin{equation}\label{eq:K0}
 K_0(x) = \int_1^\infty \mathrm{d}t\; e^{-xt}
 \left( \frac{1}{t^3} + \frac{1}{2t^5} \right) \sqrt{t^2 - 1} .
\end{equation}
Because the potential is confined to a few $\lambdabar$ around the
nucleus, it couples directly mainly to orbitals that have
appreciable amplitude there, specifically the $s$ and $p_{1/2}$ subshells, and stabilizes
them. Every other orbital is affected mostly indirectly, through the change in
screening when the core orbitals respond. The higher-order Wichmann--Kroll
and K\"{a}ll\'{e}n--Sabry vacuum-polarization terms are not
included.\cite{Wichmann56_843,Sabry55_17,Flynn25_042811}

\subsection{Self-energy: The Pyykk\"{o}--Zhao Potential}
\label{sec:se}

The self-energy, the emission and reabsorption of a virtual photon by the
bound electron, is nonlocal and has no
exact local form. The effective potentials are therefore constructed to reproduce
accurate bound-state QED results for hydrogen-like
ions.\cite{Saue22_164101} We use the Pyykk\"{o}--Zhao local model
potential,\cite{PyykkoZhao03_1469}
\begin{equation}\label{eq:pz}
 V_\mathrm{SE}(r) = B(Z)\, e^{-\beta(Z) r^2},
\end{equation}
where $Z$ is the nuclear charge and $B$ and $\beta$ are quadratic in $Z$,
$B(Z) = b_0 + b_1 Z + b_2 Z^2$ and
$\beta(Z) = \beta_0 + \beta_1 Z + \beta_2 Z^2$. The six coefficients were fitted to
self-energy corrections to the $2s$ levels of hydrogen-like ions and to the
hyperfine splitting of hydrogen-like and lithium-like ions,\cite{PyykkoZhao03_1469,Saue22_164101}
and the authors restrict the validated domain of the potential to $ns$
electrons with $n > 1$, noting that ``direct effects on non-$s$ states''
need further work.\cite{PyykkoZhao03_1469} With $B > 0$ the potential is
repulsive, opposite in sign to the Uehling potential. Because it is a single
nucleus-centered Gaussian, it also couples directly mainly to the $s$ and
$p_{1/2}$ subshells.

The Pyykk\"o–Zhao and Flambaum–Ginges potentials agree on $s$ orbitals but show disagreement for orbitals of higher angular momentum. In variational four-component calculations on Au, Sunaga, Salman, and Saue found their $s_{1/2}$ shifts to agree to within 5\%, but the $np_{3/2}$ shifts to have opposite sign and the $d$ and $f$ shifts to differ by up to a factor of 1.6.\cite{Saue22_164101} Janke \emph{et al.}\ reach similar conclusions in two-component ZORA calculations.\cite{Janke25_104111} The difference arises because the Flambaum--Ginges potential, unlike a single nucleus-centered Gaussian, also acts directly on orbitals with comparatively small amplitude at the nucleus.\cite{Flambaum05_052115,Saue22_164101}

\subsection{Orbital Energy Shifts and Orbital Relaxation}
\label{sec:shifts}

Let X be any set of the operators above added to $\hat{H}_\mathrm{DC}$. The
self-consistent shift of subshell $n\kappa$ under X is
\begin{equation}\label{eq:shift}
 \dsc_{n\kappa}(\mathrm{X}) = \varepsilon_{n\kappa}^{(\mathrm{X})}
 - \varepsilon_{n\kappa}^{(\mathrm{DC})},
\end{equation}
where $\varepsilon_{n\kappa}^{(\mathrm{X})}$ is the orbital energy from a
self-consistent field (SCF) calculation in which X is included in the Fock
operator. The principal quantum number is given by $n$ and
$\kappa = (l - j)(2j + 1)$ 
with $l$ the orbital angular momentum of the large component and $j$ the
total angular momentum. The pair $n\kappa$ thus labels one
of the 24 occupied subshells $nl_j$, the notation used in the rest of this
work. The gauge-term shift is
$\dsc(\mathrm{gauge}) = \dsc(\mathrm{Breit}) - \dsc(\mathrm{Gaunt})$.

If instead the orbitals are not re-optimized, one obtains the frozen-orbital
shift
\begin{equation}\label{eq:frozen}
 \dfz_{n\kappa}(\mathrm{X}) = F^{(\mathrm{DC+X})}_{pp}
 - \varepsilon^{(\mathrm{DC})}_{n\kappa} ,
\end{equation}
where $p$ is a composite spinor index for a spinor belonging to subshell $n\kappa$, and
$F^{(\mathrm{DC+X})}_{pp}$ are the diagonal elements (in the basis of the DC orbitals) of the Fock matrix of $\hat{H}_\mathrm{DC} + \text{X}$
constructed from the converged DC density matrix. The frozen-orbital gauge-term shift is
$\dfz(\mathrm{gauge}) = \dfz(\mathrm{Breit}) - \dfz(\mathrm{Gaunt})$.
Quantities labelled by $n\kappa$, such as those in
\cref{eq:shift,eq:frozen}, are averages over the spinors of the subshell,
which are degenerate to numerical precision for the closed-shell atoms and
ions studied here.

It is important to note that $\dfz$ is not the 
full first-order
change in an orbital energy formally entailed by perturbation theory, which also contains the
response
of the mean
field.\cite{Janke25_104111} We therefore refer to it as the \emph{frozen-orbital}
shift rather than the first-order shift or a perturbative shift.

%===============================================================================
\section{Results and Discussion}
%===============================================================================

\subsection{Computational Details}
\label{sec:compdet}

All calculations were performed at the Dirac--Hartree--Fock level with a development version of the
\texttt{ChronusQ} package.\cite{Li20_e1436,Li26_046102} We use Dyall's all-electron
triple-zeta basis throughout.\cite{Dyall06_441,Dyall09_12638,Dyall07_491} Restricted kinetic
balance is used in the four-component formalism.\cite{Dyall90_25,Nieuwpoort91_131} Nuclei
are described by a Gaussian
charge distribution with the standard mass-dependent
exponents.\cite{Dyall97_207}
The SCF convergence criteria were $10^{-10}\,\Eh$ on the energy, $10^{-10}$ on the
RMS change in the density matrix, and $10^{-8}$ on its largest element. The value of the speed of
light used in all calculations in atomic units was $c=137.035999177$.

For each member we run eight self-consistent calculations: Dirac--Coulomb
(DC); DC plus one of Uehling, self-energy, Gaunt, or Breit; DC plus Uehling and
self-energy together; and Gaunt or Breit plus both QED potentials. Four further calculations
per member add Gaunt, Breit, Uehling, or self-energy to a Fock matrix built
once from the converged DC density, which gives the frozen-orbital shifts of
\cref{eq:frozen}.

The Uehling potential of \cref{eq:uehling_1D} is evaluated for a Gaussian
nuclear charge distribution using the expansions of Fullerton and
Rinker\cite{FullertonRinker76_1283} to calculate the $K_0$ integrals. We note that other expansions, such as those due to Huang,\cite{Huang76_1311} are also available. The Pyykk\"{o}--Zhao
coefficients are
those of the original fit.\cite{PyykkoZhao03_1469}
The matrix elements of both potentials in
\cref{eq:hLL,eq:hSSs,eq:hSSJ} are evaluated by numerical quadrature on an
atom-centered grid with Becke partitioning.\cite{Becke88_2547,Murray93_997}
The radial grids\cite{Murray93_997} are modified using a scaling factor tuned to generate grid points
with adequate coverage for potentials very close to the nucleus. We note that the implementation is completely general and can be applied to molecules, as demonstrated with its application to TlBr.

The triple-zeta basis is adequate for the gauge shifts reported here. Hoyer and
Holen tabulate the Dirac--Coulomb, Gaunt, and Breit orbital energies of every
Rn subshell to $1$~meV, with the all-electron Dyall
quadruple-zeta basis.\cite{Hoyer26_5851} Our self-consistent Gaunt, gauge and
Breit shifts agree with theirs to within $\approx 0.001$~eV, the precision of their
tables, for all 24 subshells. The absolute
Dirac--Coulomb orbital energies differ by $0.002$--$0.004$~eV (at most
0.03\% of the orbital energy), which is largely irrelevant here because for orbitals we examine only
energy shifts.

Finally, we note that our results fall into two classes. Comparisons with the Uehling potential alone, and comparisons on the $s$ subshells, should not substantially depend on the choice of self-energy model. Statements about the self-energy on the $p$, $d$ and $f$ subshells, namely its sign, the role of core relaxation in its shift, and the ratio of the gauge term to the total QED shift, are properties of the Pyykk\"o–Zhao potential and may differ with the Flambaum--Ginges potential. Throughout, ``self-energy'' denotes the Pyykk\"o–Zhao potential, and we do not repeat this qualification for each result.

\subsection{Relative Magnitudes of Breit and QED Corrections}
\label{sec:total}

\begin{figure}[t]
    \centering
    \includegraphics[width=\linewidth]{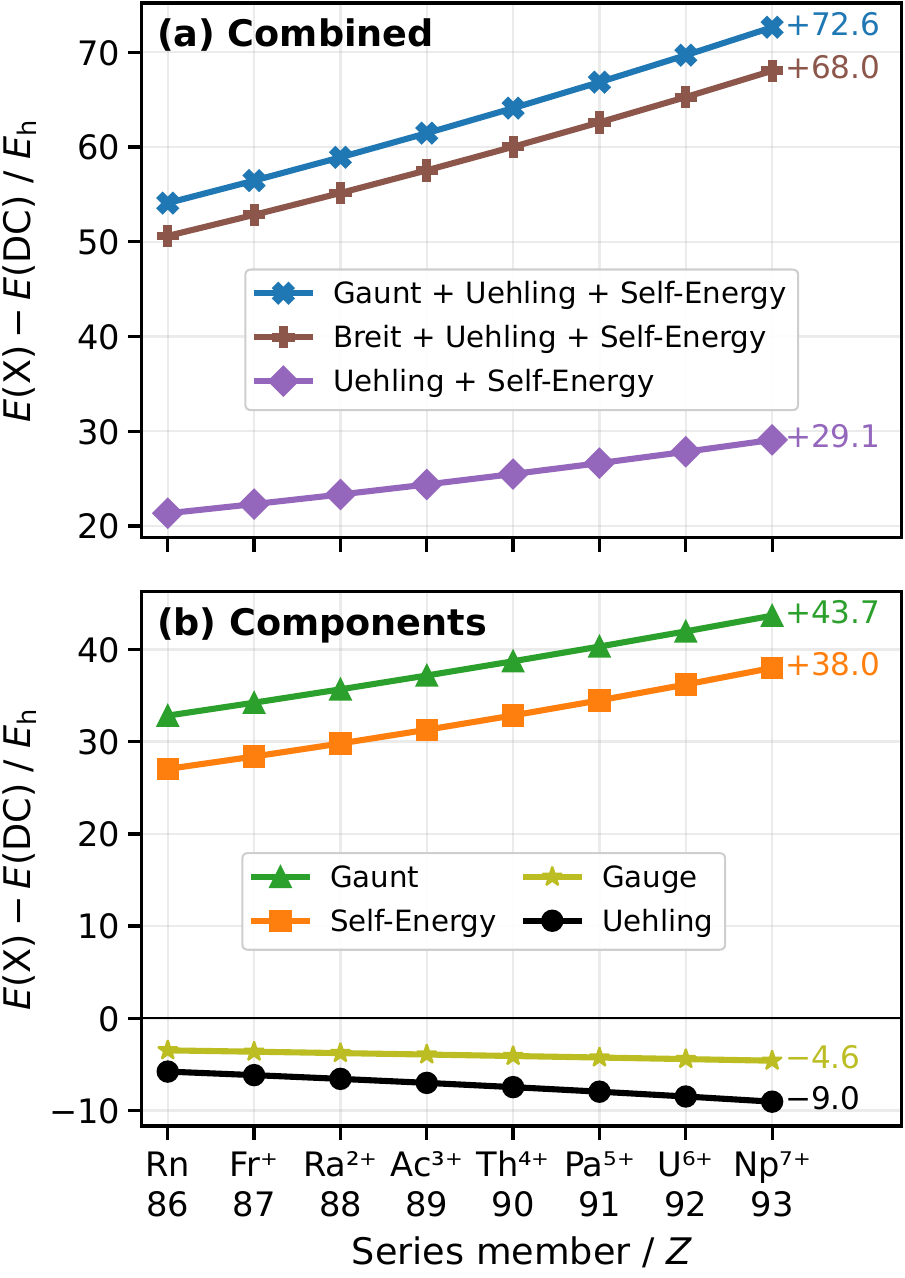}
    \captionsetup{width=\linewidth}
    \caption{Self-consistent shift in the total electronic energy relative to
    the Dirac--Coulomb baseline across the radon-like isoelectronic series,
    for each operator set X added to the Dirac--Coulomb Hamiltonian. Panel (a) shows calculations run with operator combinations while panel (b) displays single-operator calculations.
    The gauge term in panel (b) is calculated according to $E(\mathrm{Breit}) - E(\mathrm{Gaunt})$. Values at
    Np$^{7+}$ ($\Eh$) are given beside the last point of each trace. All eight members carry 86
    electrons.}
    \label{fig:total}
\end{figure}

\Cref{fig:total}a presents the total-energy shifts obtained from variational calculations using different relativistic Hamiltonians, with the Dirac--Coulomb Hamiltonian as the reference. For the series considered here, corrections beyond the Dirac--Coulomb Hamiltonian, including the Breit interaction, Uehling potential, and self-energy contributions, produce total-energy shifts of roughly 50--70 $\Eh$, with their magnitudes increasing linearly with atomic number.

\Cref{fig:total}b quantifies the contributions of individual relativistic corrections to the total energy. From Rn to Np$^{7+}$, the Uehling potential lowers the total energy by $5.77$–$9.03\,\Eh$, whereas the self-energy correction raises it by $27.04$–$38.00\,\Eh$. The combined radiative corrections, evaluated self-consistently, therefore yield a net increase of $21.34$–$29.09\,\Eh$. In comparison, the gauge term lowers the total energy by $3.47$–$4.59\,\Eh$, while the Gaunt interaction produces a larger positive contribution of $32.82$–$43.66\,\Eh$.

Importantly, the gauge contribution is substantial, amounting to 60\% of the Uehling correction for Rn and 51\% for Np$^{7+}$, and consistently offsets approximately 16\% of the net radiative correction across the series. To put this magnitude into perspective, the higher-order vacuum polarization contributions from the Wichmann--Kroll and K\"{a}ll\'{e}n--Sabry terms, which are neglected in the present work, amount to only about 6\% of the Uehling correction even at $Z=119$.\cite{Flynn25_042811} Thus, the gauge contribution is approximately an order of magnitude larger than these higher-order vacuum polarization effects, highlighting its quantitative importance in accurate relativistic total-energy calculations.

\subsection{Orbital-Resolved Gauge and QED Shifts}
\label{sec:main}

Each self-consistent orbital-energy shift of \cref{eq:shift} can be understood in terms of two
quantities: The frozen-orbital shift, $\dfz$ of \cref{eq:frozen}, which we describe as the
direct effect of the operator on the Dirac--Coulomb orbital, and
the difference between the self-consistent and frozen-orbital shifts, which arises
when the orbitals respond to the operator and the screening of the nucleus
changes. The latter is the result of orbital relaxation. The frozen-orbital shifts of every member are
given in Tables S20--S27 of the Supporting Information. We first describe the direct
shifts and then what changes when the orbitals relax.

\begin{figure*}[t]
    \centering
    \includegraphics[width=\linewidth]{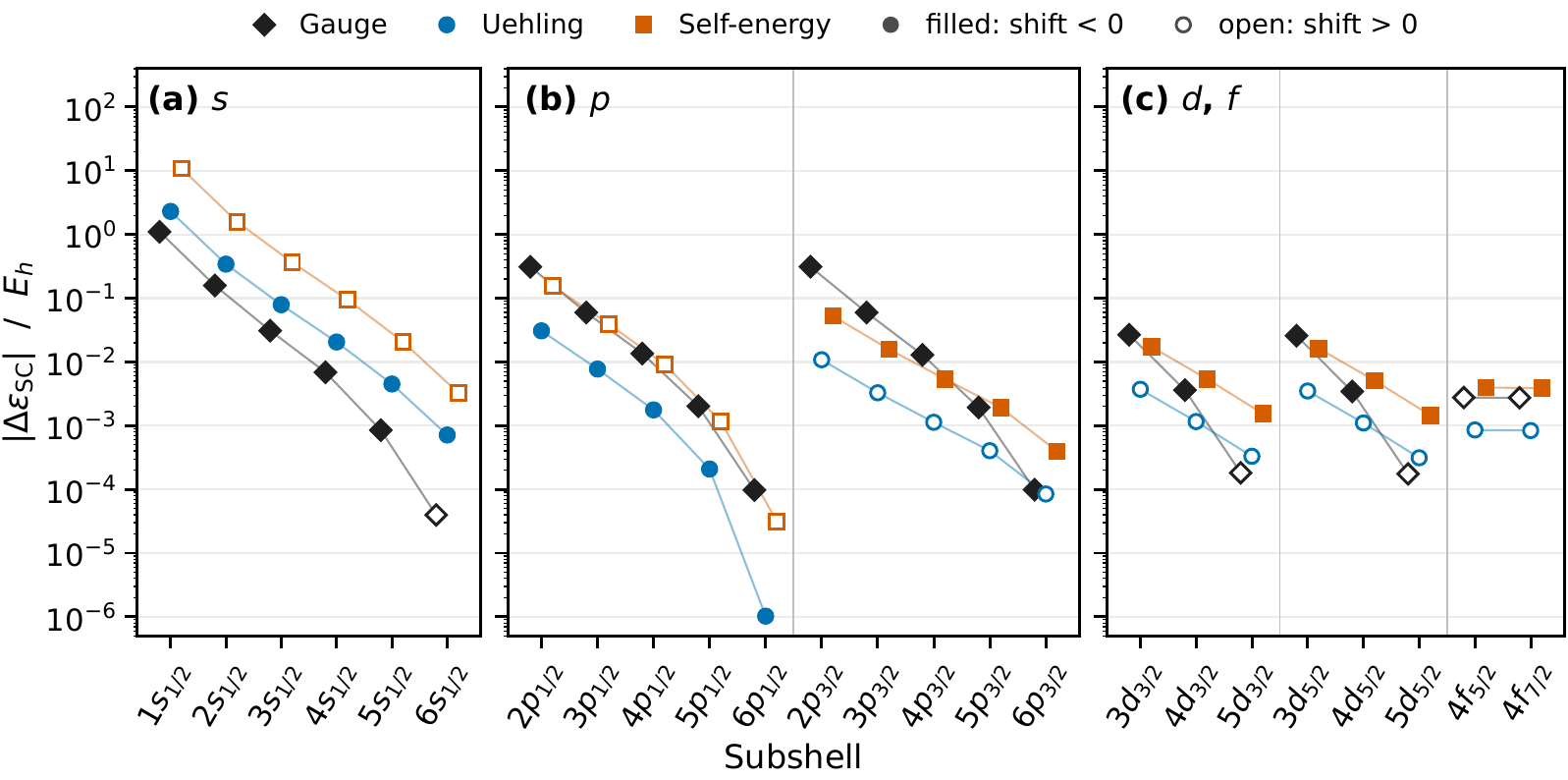}
    \captionsetup{width=\linewidth}
    \caption{Magnitude of the self-consistent orbital-energy shift at Rn
    for the gauge term (Breit $-$ Gaunt), the Uehling potential, and the
    Pyykk\"{o}--Zhao self-energy potential. Panels (a), (b) and (c) are
    grouped by subshells $s$, $p$, and combined $d, f$, respectively.
    Filled markers are stabilizing shifts, open markers destabilizing. The
    ratio of the gauge-term shift to the combined QED shift is given for
    every member in \cref{fig:heatmap}.}
    \label{fig:gaugeqed}
\end{figure*}

\begin{figure}[ht!]
    \centering
    \includegraphics[width=\linewidth]{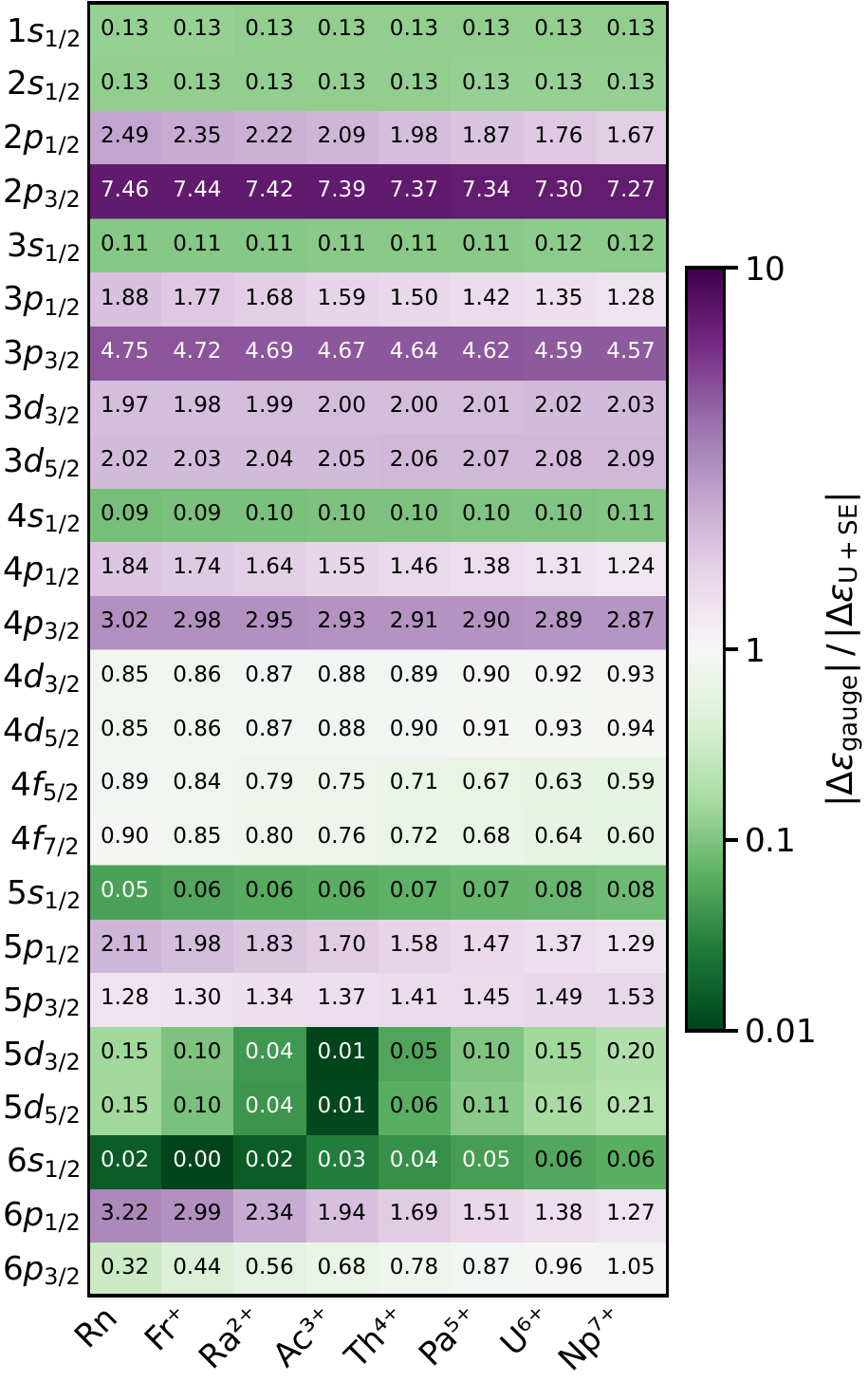}
    \captionsetup{width=\linewidth}
    \caption{Ratio of the self-consistent gauge-term shift to the combined
    Uehling and Pyykk\"{o}--Zhao self-energy shift,
    $|\dsc(\mathrm{gauge})|/|\dsc(\mathrm{U{+}SE})|$,
    for every occupied subshell (rows) and every member of the series
    (columns). Purple: the gauge term is larger. Green: the radiative
    corrections are larger. Logarithmic color scale centered on 1.}
    \label{fig:heatmap}
\end{figure}

\begin{figure}[ht!]
    \centering
    \includegraphics[width=\linewidth]{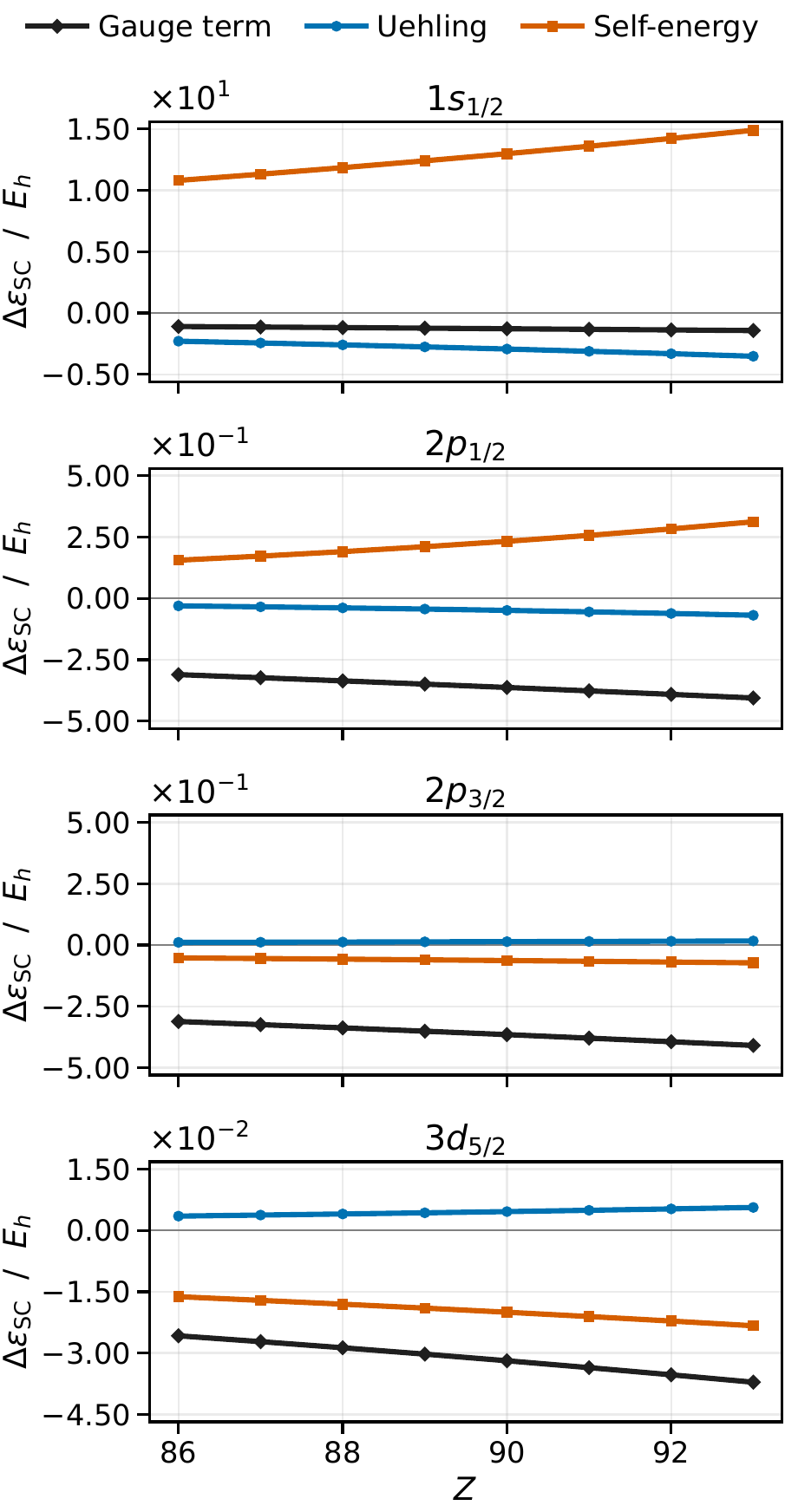}
    \captionsetup{width=\linewidth}
    \caption{Self-consistent orbital-energy shifts against nuclear charge for
    the gauge term, the Uehling potential, and the
    Pyykk\"{o}--Zhao self-energy potential for four representative subshells
    $1s_{1/2}$, $2p_{1/2}$, $2p_{3/2}$, and $3d_{5/2}$.}
    \label{fig:ztrends}
\end{figure}

\subsubsection{Direct Shifts}

With frozen orbitals each correction has a consistent sign on every subshell of
every member (Tables S20--S27). The gauge term and the Uehling potential
stabilize, and the self-energy destabilizes the orbital energy. The corrections differ in where
their direct shift is large. The Uehling and self-energy potentials are
confined to a few $\lambdabar$ around the nucleus
(\cref{sec:uehling,sec:se}), and therefore act mainly on the $s_{1/2}$ and
$p_{1/2}$ subshells, which have appreciable amplitude in this region. The direct QED shifts are substantially smaller for the $p_{3/2}$ orbitals than for their $p_{1/2}$ partners, decrease further for the $d$ orbitals, and become too small to resolve within the numerical precision of the calculations for the $4f$ orbitals. The gauge term is a two-electron
operator and is not confined in the same way. Its direct shift varies much less across the subshells of a given shell than the QED shifts do, and generally decreases from the core
outward.

\subsubsection{Self-consistent Shifts}
\label{sec:relax}

The signs of the self-consistent shifts in \cref{fig:gaugeqed} are affected by the relative
direct and relaxation contributions to the shift. On almost every subshell relaxation acts against the direct shift, so that the signs of the shifts are set by whichever of the two contributions is larger. For the QED potentials, the direct shift is the larger on the $s_{1/2}$ and
$p_{1/2}$ subshells, and on the $s_{1/2}$ subshells relaxation changes it by only a few
percent. On every other subshell relaxation is the larger effect. The Uehling and
self-energy shifts therefore keep their direct signs on $s_{1/2}$ and
$p_{1/2}$ and take the opposite signs on $p_{3/2}$, $d$, and $f$. For the gauge term, relaxation cancels a
growing share of the direct shift from the core outward. It outweighs the direct shift on
$4f$, whose direct shift is about half that of $4d$ while its relaxation is nearly the same,
so that the $4f$ gauge shift is destabilizing for every member. On $5d$ and $6s$ the two
nearly cancel, and relaxation is the larger only at the start of the series. The gauge shift is stabilizing on all other subshells.

The same picture applies to the relative size of the gauge and QED shifts
(\cref{fig:heatmap}). On the $s$ subshells, where the QED potentials act most
strongly, the total QED shift is much larger, and the gauge term offsets
about 13\% of it in the core, although it is nearly half of the Uehling shift
alone. On $p_{1/2}$ the gauge shift is about twice the total QED shift and of
opposite sign, so that it reverses the sign of the net shift. On $p_{3/2}$,
$d$, and $f$, where the QED shifts come from relaxation, the gauge term is larger than
the total QED shift on 2$p_{3/2}$--5$p_{3/2}$ and 3$d$, and
smaller on 4$d$, 4$f$, 5$d$, and 6$p_{3/2}$. Toward the
valence both shifts decrease, but the gauge shift decreases faster, because
relaxation cancels a growing share of it, whereas the QED shift, itself a
screening response, decreases more slowly. The ratio therefore falls from
about two on $3d$ to just below one on $4d$, and is smallest on $5d$ and (at the start of the series)
$6p_{3/2}$, where relaxation has removed most of the gauge shift.

Along the series (\cref{fig:ztrends}) all shifts grow in magnitude as the
orbitals contract with increasing $Z$. The ratio is nearly constant on $1s$ and $2s$
and decreases on $p_{1/2}$ as a function of $Z$, because the QED shift grows faster. In the valence it
becomes larger, because relaxation removes a shrinking share of the gauge shift as $Z$
increases.

\subsubsection{Frozen-orbital and Self-consistent Ratios}

\begin{figure}[ht!]
    \centering
    \includegraphics[width=\linewidth]{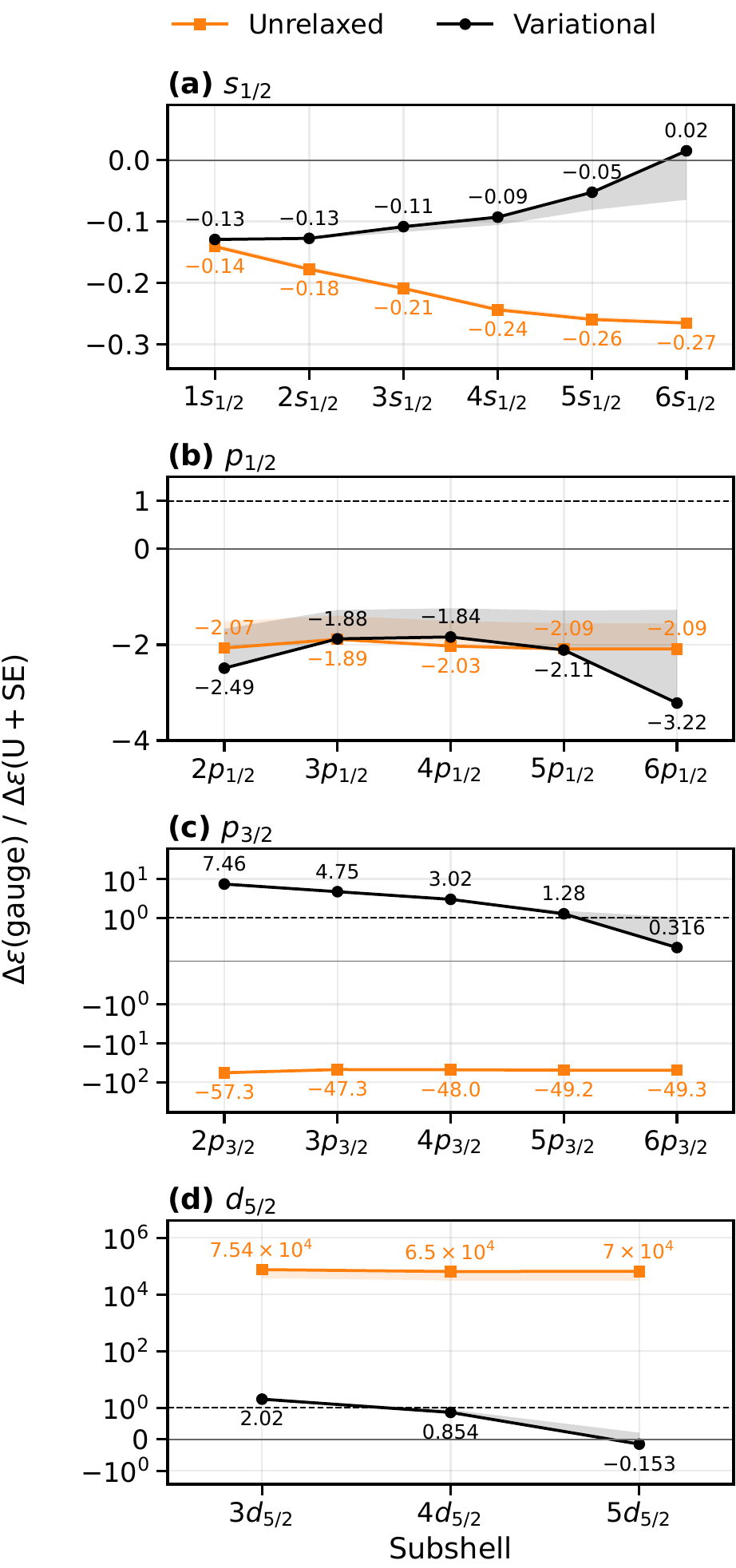}
    \captionsetup{width=\linewidth}
    \caption{Signed ratio of the gauge-term shift to the total radiative QED shift on the (a) $s_{1/2}$, (b) $p_{1/2}$, (c) $p_{3/2}$, and (d) $d_{5/2}$
    subshells, with frozen DC orbitals (unrelaxed, orange) and
    self-consistently (variational, black). Lines at Rn, bands over
    Rn--Np$^{7+}$. Values are labelled at Rn. Horizontal dashed lines indicate a ratio of 1.}
    \label{fig:ratio_spd}
\end{figure}

Because relaxation changes the gauge and QED shifts by different fractions,
frozen orbitals and relaxed orbitals can give different relative sizes (\cref{fig:ratio_spd}).
On the $s$ subshells (\cref{fig:ratio_spd}a) the frozen-orbital ratio grows in
magnitude toward the valence, whereas the self-consistent
ratio falls nearly to zero, because relaxation cancels the gauge shift but
leaves the QED shift almost unchanged. The two agree only at $1s$. On
$p_{1/2}$ (\cref{fig:ratio_spd}b) the two ratios nearly coincide on
$3p_{1/2}$--$5p_{1/2}$. This does not mean that relaxation is negligible, but
that it removes similar fractions of the gauge and QED shifts. On $p_{3/2}$
and $d_{5/2}$ (\cref{fig:ratio_spd}c,d) the frozen-orbital QED shift is small,
so the frozen-orbital ratio is large. On $p_{3/2}$ the frozen-orbital QED shift also has the opposite sign to the self-consistent one, so the ratio changes sign when the orbitals relax. Frozen and self-consistent orbitals therefore agree on the relative size of the gauge and QED shifts only on $1s$ and on $3p_{1/2}$--$5p_{1/2}$. 

\subsection{Effects of Orbital Relaxation on Fine-Structure Splittings}
\label{sec:so}

\begin{figure*}[t]
    \centering
    \includegraphics[width=\linewidth]{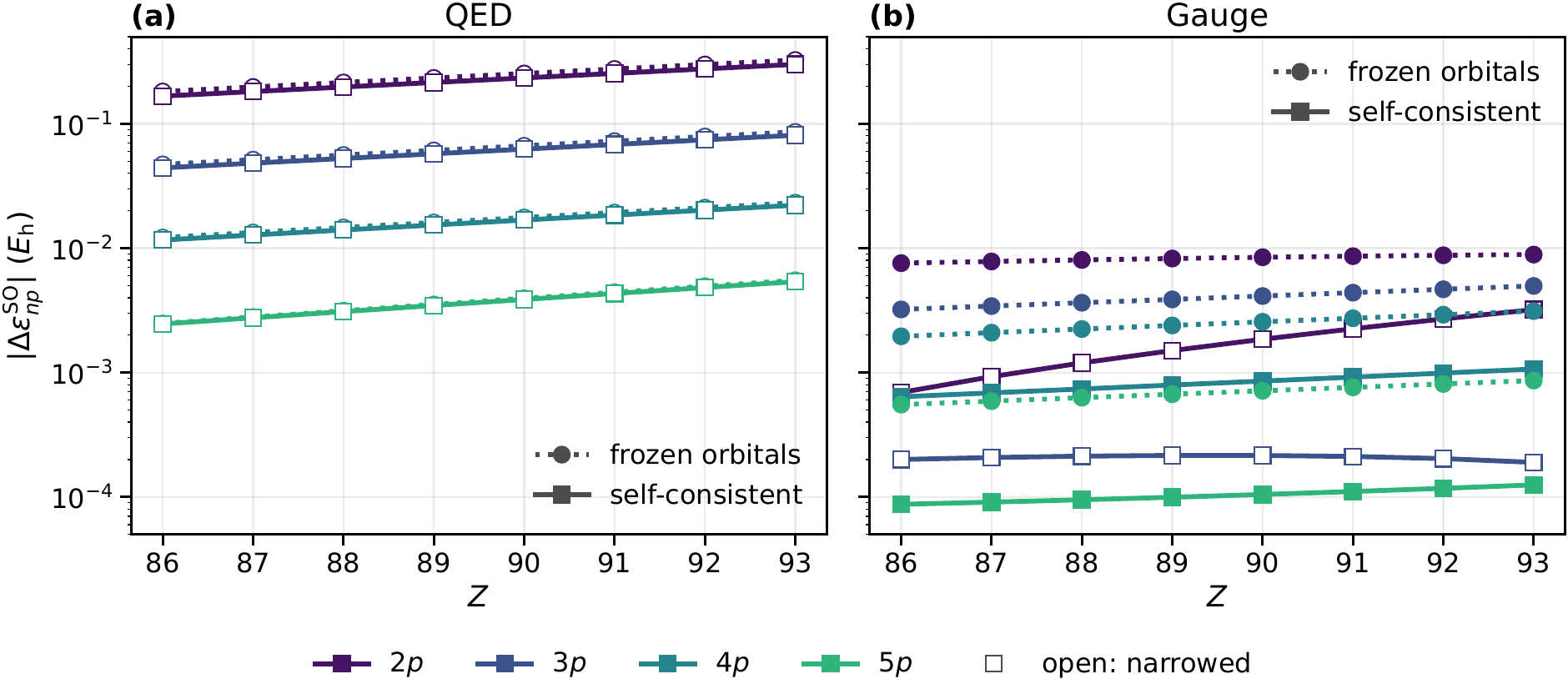}
    \captionsetup{width=\linewidth}
    \caption{Magnitude of the change in the $np$ spin--orbit splitting,
    $|\Delta\varepsilon^{\mathrm{SO}}_{np}|$, for $n = 2$--5 against nuclear
    charge with frozen DC orbitals (dotted) and self-consistently (solid).
    Panel (a) shows the QED total, while panel (b) shows the gauge term. Open
    markers indicate that the correction narrows the splitting
    ($\Delta\varepsilon^{\mathrm{SO}}_{np} < 0$), filled markers that it widens it.}
    \label{fig:so}
\end{figure*}

\Cref{fig:so} compares frozen-orbital and self-consistent gauge and QED corrections on shell-resolved fine-structure splittings arising from spin--orbit coupling. We characterize each correction X by the change it produces in the spin--orbit splitting $\varepsilon_{np_{3/2}} - \varepsilon_{np_{1/2}}$. Because every shift in \cref{eq:shift} is a difference between two calculations, this change is simply the difference between the shifts of the two partners,
\begin{equation}\label{eq:so_change}
 \Delta\varepsilon^{\mathrm{SO}}_{np} = \dsc_{np_{3/2}}(\mathrm{X}) - \dsc_{np_{1/2}}(\mathrm{X}) .
\end{equation}
The frozen-orbital $\Delta\varepsilon^{\mathrm{SO}}_{np}$ shifts are calculated using $\dfz$ (\cref{eq:frozen}) in the place of $\dsc$.
A negative $\Delta\varepsilon^{\mathrm{SO}}_{np}$ means that the correction narrows the splitting, and a positive one that it widens it. The change in the splitting therefore measures how unequally a correction acts on the two spin–orbit partners.

In the frozen-orbital treatment, the combined QED corrections shift the $np_{1/2}$ and $np_{3/2}$ orbital energies in the same direction, but the magnitude of the shift is substantially larger for $np_{1/2}$ due to its appreciable electron density near the nucleus. This differential response reduces the energy separation between the two spin--orbit components, resulting in a 0.18\%--0.25\% decrease in the fine-structure splitting compared to those from Dirac--Coulomb calculations.

In contrast, the self-consistent treatment produces energy shifts of opposite signs for the two spin--orbit partners, a trend consistently observed across all $p$ shells and atomic species considered here. The $np_{1/2}$ orbital retains the sign of its direct QED correction, whereas the $np_{3/2}$ orbital is dominated by the relaxation arising from the effective core screening, which produces an energy shift opposite in sign to the direct correction (see \cref{sec:relax}). The self-consistent treatment increases the fine-structure splitting by 0.002\%--0.017\% compared with the frozen-orbital
treatment.

The gauge contribution exhibits a different behavior. As shown in \cref{fig:so}b, the frozen-orbital gauge correction increases the fine-structure splitting of the $np$ manifold, with the magnitude of this increase growing only slowly across the atomic species considered here. However, the gauge-induced change in fine-structure splitting is consistently much smaller than the corresponding QED-induced correction seen in \cref{fig:so}a, often by at least an order of magnitude. When the orbitals relax, the gauge-induced change in the fine-structure splitting is much smaller than with frozen orbitals.

The fine-structure splittings reveal a different picture of the relative importance of gauge and QED corrections compared with the orbital energies. Although the self-consistent gauge contribution to the $p$ orbital energies exceeds the combined QED correction on the 2$p$--5$p$ subshells for every member of the series (\cref{sec:main}), it shifts the two spin--orbit partners by nearly equal amounts, resulting in only a small change in their energy separation. In contrast, the QED corrections shift the two partners in opposite directions, and produce a much larger change in the fine-structure splitting. Consequently, the self-consistent gauge contribution to the $np$ fine-structure splittings is at least an order of magnitude smaller than the combined QED contribution. These results demonstrate that the gauge interaction primarily affects the absolute positions of the $p$ energy levels, whereas QED corrections play a dominant role in modifying their fine-structure splittings.

\subsection{Equilibrium Bond Length of TlBr}
\label{sec:tlbr}

\begin{table}[t]
\centering
\caption{Equilibrium bond length $r_e$ of TlBr and its shift $\Delta r_e$
relative to Dirac--Coulomb for each Hamiltonian, followed by differences
between Hamiltonians. All values in pm. Shifts are formed from the unrounded
$r_e$ and may therefore differ in the last digit from differences of the
rounded $r_e$.}
\label{tab:tlbr}
\begin{tabular}{l S[table-format=3.2] S[table-format=+1.2, retain-explicit-plus]}
\toprule
Hamiltonian & {$r_e$} & {$\Delta r_e$} \\
\midrule
DC         & 267.03 & {--}  \\
DC+U       & 267.05 & +0.02 \\
DC+SE      & 266.93 & -0.10 \\
DC+U+SE    & 266.95 & -0.08 \\
DC+G       & 267.31 & +0.28 \\
DC+G+U+SE  & 267.23 & +0.20 \\
DC+B       & 267.26 & +0.24 \\
DC+B+U+SE  & 267.18 & +0.15 \\
\midrule
\multicolumn{2}{l}{DC+B $-$ DC+G}              & -0.05 \\
\multicolumn{2}{l}{DC+B+U+SE $-$ DC+G+U+SE}    & -0.05 \\
\bottomrule
\end{tabular}
\end{table}

\begin{table}[t]
\centering
\caption{Change in the expectation values of the Uehling and self-energy
potentials placed on Tl only and on Br only, evaluated on the Dirac--Coulomb
density of TlBr at each bond length $R$, relative to $R = 2.5682$~\AA{}, in
$\mu\Eh$. At $2.5682$~\AA{} the Uehling values are $-4.1730184$ (Tl) and
$-0.0969151$~$\Eh$ (Br), and the self-energy values $21.2845770$ (Tl) and
$1.1524541$~$\Eh$ (Br). The last row is the change from the neutral Tl atom
to the free Tl$^+$ ion.}
\label{tab:tlbr_vatom}
\begin{tabular}{S[table-format=1.4] S[table-format=+2.1, retain-explicit-plus] S[table-format=+1.1, retain-explicit-plus] S[table-format=+2.1, retain-explicit-plus] S[table-format=+1.1, retain-explicit-plus]}
\toprule
 & \multicolumn{2}{c}{Uehling} & \multicolumn{2}{c}{Self-energy} \\
\cmidrule(lr){2-3} \cmidrule(lr){4-5}
{$R$ (\AA)} & {Tl} & {Br} & {Tl} & {Br} \\
\midrule
2.5682 & {--} & {--} & {--} & {--} \\
2.6182 & -2.8 & 0.0 & +14.0 & -0.4 \\
2.6682 & -5.5 & +0.1 & +27.5 & -0.8 \\
2.7182 & -8.1 & +0.1 & +40.5 & -1.3 \\
2.7682 & -10.6 & +0.1 & +52.9 & -1.6 \\
\midrule
\multicolumn{5}{l}{\textit{Free atoms}} \\
{Tl $\to$ Tl$^+$} & {$-47.6$} & {--} & {$+232.5$} & {--} \\
\bottomrule
\end{tabular}
\end{table}

To see how the corrections compare on a molecular property, we computed the
equilibrium bond length of TlBr with the same basis, nuclear model, and
convergence criteria as for the atoms (\cref{sec:compdet}). The thallium halides are also systems in which QED effects on NMR shieldings have been investigated.\cite{Aucar24_064307,Berger26_09399} The Uehling and self-energy potentials are applied on
both atoms. For each of the eight Hamiltonians used, the
self-consistent energy was computed at five bond lengths from $2.5682$ to
$2.7682$~\AA{} in steps of $0.05$~\AA, and $r_e$ was taken as the minimum of a
quartic polynomial in $1/R$ through these points. The results are given in
\cref{tab:tlbr}, rounded half up to $0.01$~pm.

The Gaunt interaction is the largest correction and lengthens the bond by
$0.28$~pm. The gauge term has the opposite sign and shortens it by $0.05$~pm,
removing about one sixth of the Gaunt lengthening. Zhang and Cheng found a
similar trend in four-component Hartree--Fock calculations on the hydrogen
halides and dihalogens of Br, I and At.\cite{Cheng22_4537} The partial
cancellation of the Gaunt lengthening by the gauge term is therefore not
particular to TlBr.

The two QED potentials also act in opposite directions. The Uehling potential
lengthens the bond ($+0.02$~pm), while the self-energy shortens it
($-0.10$~pm), so that the total QED shift is a shortening of $0.08$~pm. These
signs follow from how the electron density near the nuclei changes with bond
length, because both potentials act only close to the nuclei. To
investigate these shifts, we evaluated the expectation values of each
potential on the
Dirac--Coulomb density at each bond length. Both expectation values grow in
magnitude as the bond is stretched. Placing each potential on one nucleus at
a time (\cref{tab:tlbr_vatom}) shows that this growth comes from Tl, where
the density near the nucleus increases with $R$. Near Br the density
decreases slightly, which changes the expectation values in the opposite
direction by only a few percent of the Tl change. The attractive Uehling
potential therefore lowers the energy more at long bond lengths and favors a
longer bond, while the repulsive self-energy raises it more and favors a
shorter bond. The self-energy change is the larger of the two, so the net
effect is a shortening.

To see why the density near the Tl nucleus changes, we use Mulliken
populations of Tl resolved by angular momentum (Tables S36 and S37 of the
Supporting Information). Each is the number of electrons
in the Tl basis functions of a given angular type ($s$, $p$, $d$ or $f$),
obtained by summing the Mulliken gross populations of those functions over all
occupied orbitals. In the closed-shell Tl$^+$ ion each occupied orbital has a
single angular type, so the populations are exactly the shell occupations,
$12$ $s$ and $24$ $p$ electrons. The deviations from these values in TlBr
therefore show how the bond redistributes the Tl density between angular
types. At the shortest bond length, Tl carries about $0.48$ additional
electrons in $p$-type functions and a deficit of about $0.07$ in $s$-type
functions relative to the free Tl$^+$ ion.

Both the loss of $s$-type population and the gain of $p$-type population
lower the density at the Tl nucleus, which is carried mainly
by the $s$ electrons. The $s$ deficit does so directly. The additional $p$
population does so indirectly, by partly screening the $6s$ density from the
nucleus,\cite{Steudel74_613} as in the free atoms, where removing the $6p$
electron from Tl lowers the Uehling expectation value and raises the
self-energy expectation value (\cref{tab:tlbr_vatom}). The effects of valence $p$ shell filling on $s$ core density are
known from M\"{o}ssbauer isomer shifts of tin compounds, where adding $5p$
electrons lowers the $5s$ density at the nucleus, and $5s$--$5p$ mixing in
stannous compounds lowers it below that of the ideal $5s^2$
ion.\cite{Flinn68_882} As the TlBr bond is stretched it becomes more ionic. The
Mulliken charge on Tl rises from $+0.54$ to $+0.59$ over the grid, while the $p$
excess decreases and the $s$ deficit shrinks (Table S36). The net result of these effects is an increase of the density near the Tl nucleus, as observed from the trends in
\cref{tab:tlbr_vatom}.

The sign of the total QED shift agrees with some earlier molecular
results. QED lengthens bonds in which the valence $s$
orbital of the heavy atom takes part, as in AuCN, AuF and
Cn$_2$,\cite{Saue22_164101,Peterson26_2571687,Hangele14_222} but was found to
shorten them slightly in some hydrides of superheavy $p$-block elements.\cite{Hangele13_044104} Sunaga,
Salman and Saue suggested that this opposite trend may arise when the bonding
valence orbitals have $p$ rather than $s$ contributions from the heavy
atom.\cite{Saue22_164101} TlBr appears to be such a case. Bonding adds $p$ rather than
$s$ population to Tl, and the resulting $p$ excess and $s$ deficit lower the
density near the Tl nucleus. Because both grow as the bond is compressed, the net
repulsive QED correction favors the shorter bond.

The total QED shift is $-0.08$~pm whether it is added to the
Dirac--Coulomb, Dirac--Coulomb--Gaunt or Dirac--Coulomb--Breit Hamiltonian, and
the gauge term is $-0.05$~pm with or without the QED potentials. Both the
gauge term and the total QED shift shorten the bond, and the gauge term amounts to more than half of the total QED shift and more than twice
the Uehling shift. For this case, a Dirac--Coulomb--Gaunt calculation with the QED potentials
thus omits a contribution to $r_e$ of the same order as the radiative
correction it includes.

%===============================================================================
\section{Conclusion}
%===============================================================================

We have added the Uehling and Pyykk\"{o}--Zhao potentials to the
\texttt{ChronusQ} program, alongside the full Breit interaction, and used
them to compare the gauge term with the radiative corrections in the
radon-like series Rn--Np$^{7+}$ and in TlBr.

On the core $s$ shells the gauge term has the same sign as the Uehling shift
and about half its size, although the self-energy makes the total QED shift
much larger. On nearly every other subshell the gauge term exceeds the
Uehling shift, and on most $p$ subshells and on $3d$ it exceeds the total
QED shift. Orbital relaxation separates the two further. The radiative shifts of
the $s$ shells barely change, and those of the $d$ and $f$ subshells come almost
entirely from the response of the core.

In TlBr the gauge term removes about one sixth of the Gaunt lengthening of
the bond, and the total QED shift shortens it. The shortening arises because
bonding adds $p$ rather than $s$ population to Tl, which lowers the density
at the Tl nucleus most at short bond lengths. The gauge term and the total
QED shift have the same sign, and the gauge term is more than half as large.

A Dirac--Coulomb--Gaunt calculation with effective QED potentials therefore
omits a contribution comparable to the radiative corrections it includes,
both for orbital energies outside the $s$ shells and for a molecular property.
Where an accuracy target requires vacuum polarization, it also requires the
full Breit interaction. The Uehling and $s$-shell comparisons should hold for either
common self-energy model; on the $p_{3/2}$, $d$ and $f$ subshells the
self-energy results are those of the Pyykk\"{o}--Zhao potential, which
differs there from the Flambaum--Ginges potential. Future work will include comparisons with the Flambaum--Ginges potential, other self-energy models, higher-order corrections, as well as the inclusion of correlation effects.

\section*{Acknowledgment}

The development of quantum field electronic structure theory is supported by the U.S. Department of Energy, Office of Science, Basic Energy Sciences, in the Computational and Theoretical Chemistry program (Grant No. DE-SC0006863). The development of the Chronus Quantum computational software is supported by the Office of Advanced Cyberinfrastructure, National Science Foundation (Grant No. OAC-2103717). The authors acknowledge the use of facilities and instrumentation supported by the U.S. National Science Foundation through the UW Molecular Engineering Materials Center (MEM-C), a Materials Research Science and Engineering Center (DMR-2308979). The authors thank Trond Saue, Kirk Peterson, and Lucas Visscher for helpful and productive discussions.

%===============================================================================
\section*{Data Availability}
%===============================================================================
The data that support the findings of this study are available within the article.

\section*{Supporting Information}
The Supporting Information tabulates the total energies and the orbital
energies of all 24 occupied subshells for each of the members in
the isoelectronic series. Frozen-orbital shifts are also reported as well as
Mulliken population information related to TlBr.

\providecommand{\latin}[1]{#1}
\makeatletter
\providecommand{\doi}
  {\begingroup\let\do\@makeother\dospecials
  \catcode`\{=1 \catcode`\}=2 \doi@aux}
\providecommand{\doi@aux}[1]{\endgroup\texttt{#1}}
\makeatother
\providecommand*\mcitethebibliography{\thebibliography}
\csname @ifundefined\endcsname{endmcitethebibliography}
  {\let\endmcitethebibliography\endthebibliography}{}

\end{document}